# Supplemental Data

# Stochastic Gene Expression

# in a Lentiviral Positive-Feedback Loop:

# HIV-1 Tat Fluctuations Drive Phenotypic Diversity

Leor S. Weinberger, John C. Burnett, Jared E. Toettcher, Adam P. Arkin, and David V. Schaffer

## Table of Contents

# 1. Integration Data
## Integration Sites of LGIT PheB Clones

| | Clone | Chrmsm | SINE | LINE | LTR | DNA | mRNA | Overlapping or adjacent Gene | telomere | dir. | start pos. | Gene Expressed? |
|---|---|---|---|---|---|---|---|---|---|---|---|---|
| | | | | | *Repeats* | | | | | | | |
| 1 | E7 | 8 | AluJo 279 | / | 1052 LTR12C ERV1 | / | P | predicted only | 0.2Mb | - | 146107834 | No Data |
| 2 | D8-2 | 10 | / | 303 L1MB4 | 1450 LTR67 ERVL | 292 MER7D MER2_type | / | LOC390009 similar to 40S ribosomal protein S26 Ê | 11Mb | - | 124506765 | No Data |
| 3 | F8-1 | 5 | F | 1kb | / | / | P | NSD1 nuclear receptor biNo Datating SET domain protein 1 | 5Mb | + | 176648028 | YES IN JURKATS |
| 4 | E7-2 | 11 | AluYd2 291 | X L1ME | 466 MLT1D MaLR | 1.5kb | P | C11orf23 NM_018312 | No | + | 68105564 | HIGHLY in Jurkats |
| 5 | F5-2 | 10 | / | 303 L1MB4 | 1450 LTR67 ERVL | 292 MER7D MER2_type | / | LOC390009 similar to 40S ribosomal protein S26 Ê | 11Mb | - | 124506765 | No Data |
| 6 | F9-1 | 6 | 1379 AluSx | 146 L1Mc5 | / | 250 Charlie7 | P | UTRN utrophin (homologous to dystrophin) | 26Mb | + | 144756507 | YES IN JURKATS |
| 7 | B6-2 | 1 | MIRb 210 | / | 817 MSTA MaLR | ÊDNAJC8 ~100bp 5' | / | ? | 28Mb | + | 28131671 | No Data |
| 8 | G9-1 | 16 | 1.25kb | 1kb | / | 500 | AP | ÊLAT1-3TM NM_031211 | No | + | 29484644 | HIGHLY in Jurkats |
| 9 | E1 (ssp) | X | 210 AluJb | X L3;Cr1 | / | Mer2_type 843bp | P | AK125301 FLJ43311 | 12Mb from | - | 72307977 | No |
| 10 | G4-1 | 8 | / | / | / | / | AP | RNF19 ring finger protein 19 | No | - | 101238123 | lymphocyte, low in lymph but Not expressed in jurkats |
| 11 | E6-1 | 10 | 1.7kb; AluSg | X; L1MB4 | 1450 LTR67 ERVL | 200bp; MER7D | AP | yeast mitotic checkpoint protein AA448112 | 11Mb | - | 124506765 | No Data |
| 12 | B10-2 | 10 | 1.7kb; AluSg | X L1MB4 | 1450 LTR67 ERVL | X MER7D | AP | yeast mitotic checkpoint protein AA448112 | 11Mb | - | 124506765 | No Data |
| 13 | D9-1 ssp | 17 | AluYC2 | L1MEc 1099 | / | / | P | ? | 1Mb | + | 80745016 | No Data |
| 14 | D10-2 | 2 | MIR 2455 | L1PA41783 | X MLT2A2 +F33ERVL | / | P | LRP1B low density lipoprotein-related protein 1B (deleted in tumors) | No | - | 141808074 | Doesn't appear expressed in Jurkats |
| 15 | E9-2 | 2 | MIR 2455 | L1PA41783 | X MLT2A2 +F33ERVL | / | P | LRP1B low density lipoprotein-related protein 1B (deleted in tumors) | No | - | 141808074 | No Data |
| 15 | C7-2 | 4 | X AluJb | F L1M4 | / | / | AP | MAEA Erythroblast macrophage protein EMP AF084928 | 1.17Mb | - | 1173246 | HIGHLY in Jurkats |
| 16 | C8-1 | 17 | ALuSx 129bp & 213bp | L1M4c 2185bp | / | X Tigger3b Mer2_type | P | ILF1 (a.k.a FOXK2) interleukin enhancer binding factor 1 | 0.5Mb | + | 78084161 | HIGHLY in Jurkats |
| 17 | C3br | 3 | MIR 165 | L2 81bp | MLT1B MaLR 411bp | / | / | / | No | - | 83116162 | No Data |
| 18 | F5br | 6 | AluSc 466bp | X L1MA8 | / | MER58A 1032bp | AP | C6orf167 | No | - | 97758716 | YES IN JURKATS |
| 19 | E9br | 17 | AluJo X | L1mB7 437 | / | / | / | / | / | + | 80520030 | No Data |
| 20 | G2br | ? | / | / | X MLT1I | / | / | integration into HERV LTR found at multiple genomic loci (i.e. integration site indistinguishable) | / | / | / | No Data |
| 21 | E4-1 | 3 | 529 AluSX | X L1ME3a | 1914 LTR16A1 ERVL | / | / | / | / | - | 152621312 | No Data |

Yellow and Blue highlighted clones/rows are duplicates or co-contamination between wells on the 96-well plate. Only 1 surrogate of each of these was used in statistical calculations. / implies no data available. Jurkat expression data was obtained from the BLAT genome browser, and from Lewinski et al. (2005). An "X" in any of the "Repeats" columns (SINE, LINE, LTR, DNA) indicates that the integration overlaps the given repeat, otherwise a numeric value indicates the distance of the integration site from the repeat. P and AP in the mRNA column denote Parallel or Anti-Parallel direction of mRNA expressed from the overlapping or adjacent human gene. A numeric value in the "telomere" column indicates the distance of the integration site from the end of the chromosome while a "No" indicates that the integration was greater than ~30Mb from the end of the chromosome. + or – in the "dir" column denotes direction of the integration relative to the human sequence and the numeric value in the "position" column indicates the start position of the integration on the given chromosome (July 2003 build of the Human Genome).

## Integration Sites of LGIT *Dim* Sorted Non-PheB Clones

| | Clone | Chrmsm | SINE | LINE | LTR | DNA | mRNA | Gene | telomere | dir. | start pos. | Gene Expressed? |
|---|---|---|---|---|---|---|---|---|---|---|---|---|
| | | | | | Repeats | | | | | | | |
| Non-bifurcating LGIT DIM clones | | | | | | (ACTIVATED, but NOT transactivated, BY TNF-alpha) | | | | | | |
| 1 | G5-2 | 16 | 1.2Kb; AluSX | 1kb; L1PA11 | / | 0.5Kb; Mer81 | AP | LAT1-3TM | No | + | 29614998 | Highly in Jurkats |
| 2 | E7-1 | 3 | 2kb AluJo | / | / | / | AP | CENTB2 | 3.7Kb | + | 196325555 | No Data |
| 3 | E5-1 | 19 | AluSq 653bp | L2 652bp | / | MER63A (AcHobo) 1.7kb | P | PPFIA3 | No | + | 54337776 | under expressed in thymus |
| 4 | D6-2 | 1 | X MIR; AluYc2 600bp | / | / | / | / | / | No | + | 170330731 | / |
| 5 | G11-2 | 17 | AluYc2 1249 | L1MEc 1099bp | / | / | / | / | 1Mb | + | 80745016 | / |
| 6 | F6-1 | 17 | 155bp AluSx & 187bp | / | / | X Tigger3b | P | ILF1 | No | + | 81169743 | Highly in Jurkats |
| Non-bifurcating LGIT DIM clones with extremely low GFP fluorescence (Genomewalker PCR aNo Data LTR PCR failures => un-infected) | | | | | | | | | | | | |
| 1 | f10-2 | No BLAT hit / No LTR seq. Attained | | | | | | | | | | |
| 2 | d7-2 | No BLAT hit / No LTR seq. Attained | | | | | | | | | | |
| 3 | c7-1 | No BLAT hit / No LTR seq. Attained | | | | | | | | | | |
| 4 | f7-2 | No BLAT hit / No LTR seq. Attained | | | | | | | | | | |
| 5 | c6-1 | No BLAT hit / No LTR seq. Attained | | | | | | | | | | |
| 6 | e6-2 | No BLAT hit / No LTR seq. Attained | | | | | | | | | | |
| 7 | E4-1 | No BLAT hit / No LTR seq. Attained | | | | | | | | | | |
| 8 | F6-2 | No BLAT hit / No LTR seq. Attained | | | | | | | | | | |

## Integration Sites of Non-PheB (LGIT Bright and LG) Jurkat Clones

| # | Clone | Chrmsm | SINE | LINE | LTR | DNA | mRNA | Gene ? | Expressed | telomere | dir. | start pos. |
|---|-------|--------|------|------|-----|-----|------|--------|-----------|----------|------|------------|
|   |       |        | (REPEATS) | | | | | | | | | |
| 1 | E8cv | 14 | AluSc 1006b D | X L1MC/D | MLT1G3 MaLR 3473bp | / | / | / | / | 9Mb | + | 96891870 |
| 2 | D6cv | 8 | MIR3 2356b P | X L1ME3A | / | / | P & AP | SGP; GeneID; GeneScan predicted (but no ESTs) | / | NO | + | 130546767 |
| 3 | D3cv | 14 | AluSc 830bp | X L1MC/D | / | / | / | / | / | 9Mb | + | 96891694 |
| 4 | C4cv | 2 | AluY 600bp | 1kb L1ME4a; L1 | / | Cheshire MER1 1635bp | / | / | / | NO | - | 98862772 |
| 5 | C4c | 2 | 500 | " | " | " | / | / | / | NO | - | 98862762 |
| 6 | C5c | 3 | X | / | / | / | P | predicted (Ecgene predicted, mRNAs exist) | / | NO | - | 143129958 |
| 7 | D9c | 10 | AluSx 2884 | / | / | / | AP | RTKN2 rhotekin-2 | / | NO | + | 63301560 |
| 8 | E11c | 13 | X | F | 750 | | AP | twnscn predicted | / | NO | + | 39787075 |
| 9 | D5c | 8 | AluJo; 843bp | L1PB2; 1830bp | LTR40a 5092bp | / | P | NRG1 neuregulin 1 isoform GGF2 NM_013962 (RefSeq gene) | NO in Jurkats | NO | + | 32056779 |
| 10 | G4c | 18 | AluJo; 700bp | L1ME Flanking | MLT1J (MaLR) 1425bp | MER20 385bp | AP | PTPN2 protein tyrosine phosphatase, non-receptor type 2 | YES in Jurkats | NO | + | 12853883 |
| 11 | B3c | 19 | AluSx 108bp | L2 1318bp | MLT1J2 (MaLR) 2049bp | / | AP | SH2D3A SH2 domain containing 3A | probably | NO | + | 6714171 |
| 12 | B6vb | 9 | X AluJo | / | / | X MER3 | AP | SARDH sarcosine dehydrogenase | NO | 6Mb | + | 131894444 |
| 13 | E3vb | 10 | 778 AluY | X L1MB7 | / | 677 MER5A1 | P | C10orf6 AI655902 | YES in Jurkats | 33Mb | + | 102384565 |
| 14 | C4vb | 5 | 500 200 | 1kb | 1.3kb | / | P | ERBB2IP erbb2 interacting protein | HIGHLY in Jurkats | NO | + | 65281011 |
| 15 | H10vb | 10 | AluY; 783bp | F; LIMB7 | / | Mer5A1; 682bp | P | C10orf6 | possibly | NO | + | 102384560 |
| 16 | D3vb | 17 | AluYC2; 1249b D | LIMEc; 1099 | / | / | | Genscan and alt-splice predictions | YES in Jurkats | NO | + | 80745016 |
| 17 | E9vb | 17 | X; AluSa | LIM4c; 1076bp | / | / | AP | NPL4 NM_017921 | | 1.5Mb | + | 80296718 |
| 18 | D7vb | 19 | X AluJo | L1ME4a 138bp | / | | AP | OLFM2 olfactomedin 2 AF131839 | HIGHLY in Jurkats | NO | + | 9839720 |

**Discussion on Integration Sites**

Interestingly, a fraction of *Dim* sorted LGIT clones (12 clones/33 LGIT *Dim* clones total) appeared to remain stable in the *Dim* region of fluorescence. This result is inconsistent with the potential model where the *Dim* region is unstable (Fig. 6, manuscript) but could be explained by a mutated, non-functional or fractured Tat TAR axis, potentially acquired during viral reverse transcription,. Indeed 6 of these 12 LGIT *Dim*-stable clones behaved similar to LG clones and could only be *activated* by TNFα, phorbol ester, or TSA incubation, i.e. they did not appear to be capable of exhibiting Tat transactivation (see below). The functional genotype of these clones is thus more equivalent to an LG clone than an LGIT clone. The complete LTRs of two *Dim* LGIT clones were sequenced and found to have large deletions (see below). We did not sequence the integrated proviral *gfp*, or *IRES-tat* genes in these clones, and it is possible that mutations in these elements could alter fluorescence or abrogate transactivation, respectively. Thus, the results remain consistent with a potential model in Fig. 6 (manuscript), where the *Dim* region is unstable (for a functional Tat transactivation loop).

The remaining six LGIT *Dim*-stable clones could not be *activated* to an increased level of GFP fluorescence by any of the above chemical perturbations. These results imply that despite having detectable GFP fluorescence, these non-activatable clones likely lacked large portions of one or both LTRs. In support of this LTR mutation-deletion hypothesis, attempts to amplify the integrated proviral LTRs in these clones failed: Genomewalker PCR bands were not distinct and sequences obtained did not register any integration hits on human genome BLAT queries despite multiple attempts at amplifying genomic DNA both upstream from the 5' LTR and downstream from the 3' LTR. Furthermore, attempts to determine the sequence of the 5' and 3' LTRs in these clones also failed, indicating either (1) very large deletions of both HIV-1 LTRs occurred, thereby eliminating homology with the PCR primers, or (2) these clones were never originally infected.

**Statistics**

Initially a multivariate analysis of variance (MANOVA) was performed on 8 clones, 4 PheB clones and 4 non-PheB clones. Based on past integration studies (Stevens and Griffith 1994; Wu et al. 2003; Mitchell et al. 2004) the MANOVA tested the following distinct hypotheses: 1) PheB clones integrate near SINEs, 2) PheB clones integrate near LINEs, 3) PheB clones integrate near HERVs 4) PheB clones integrate in intergenic regions or 5) PheB clones integrate near telomeric regions. Transcriptional interference (or RNAPII "trainwrecking") was also considered but lack of within-gene integrations made this hypothesis difficult to formally evaluate. MANOVA is known to correct for multiple hypotheses tested. The MANOVA yielded only 1 significant P value for integration within 1 kb of a HERV LTR (P=0.051), in all other cases the null hypothesis was accepted.

An additional 45 clones (PheB and non-PheB) were collected and examined for HERV-proximate integrations. Chi-square and binomial distribution testing were performed in Microsoft Excel, and a web-based calculator (http://www.matforsk.no/ola/index.html) was used for Fisher's exact test. Chi-square test is not considered accurate for small sample sizes (N < 50) and a Yates estimate correlation is typically used to correct when small sample sizes are compared. Fisher's exact test does not suffer from inaccuracies at small sample sizes. Below is an example of the contingency table and Fisher's exact test:

**Contingency Table**

|      |     | HERV | | |
|------|-----|-----|-----|-------|
|      |     | Yes | No  | Total |
| **PheB** | Yes | 7 | 10 | 17 |
|      | No  | 1 | 17 | 18 |

**Fisher's Exact Test**

TABLE = [ 7 , 10 , 1 , 17 ]
Left   : p-value = 0.9989671063085966
Right  : p-value = 0.0159065628476089902
**2-Tail : p-value = 0.0177657714921346**

**1.5% Agarose Gel of GenomeWalker PCR of PheB and Non-PheB Clones**

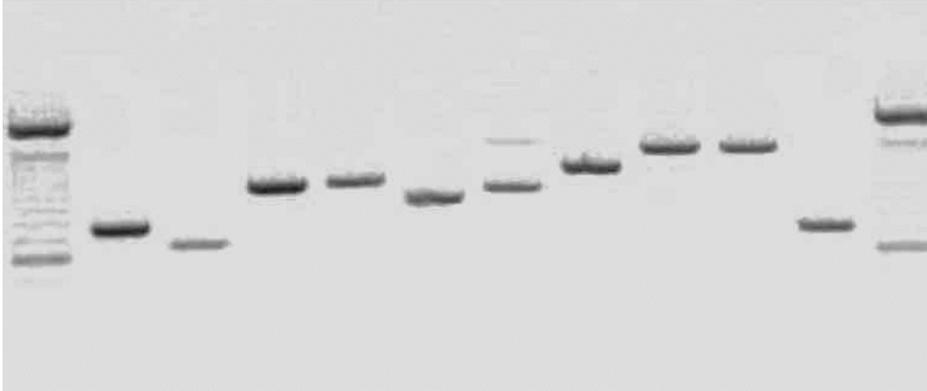

Lanes 1 and 12 are 0.5kB ladders, where the bottom-most dark band corresponds to 500bp.
PCR in Lane 7 was conducted using primers internal to the LTR and produced a single sequence of >1000 bp on the chromatograph. Much of this sequence was homologous with HIV-1; thus the extra band is consistent with PCR amplification from the 5'LTR on through the entire LGIT insert, to a *Stu* I site in the human genome just downstream of the 3' LTR (i.e. amplification in the sense direction from the 5'LTR). The higher molecular weight band in lane 7 (~1.5 kb) appears to contain significantly less DNA, also consistent with PCR of longer sequences (i.e. the entire LGIT insert from the 5'LTR to the genomic sequence flanking the 3' LTR) being less efficient than PCR of short sequences (i.e. the 3'LTR + flanking genomic sequence).

**Representative Sequence Chromatograph**

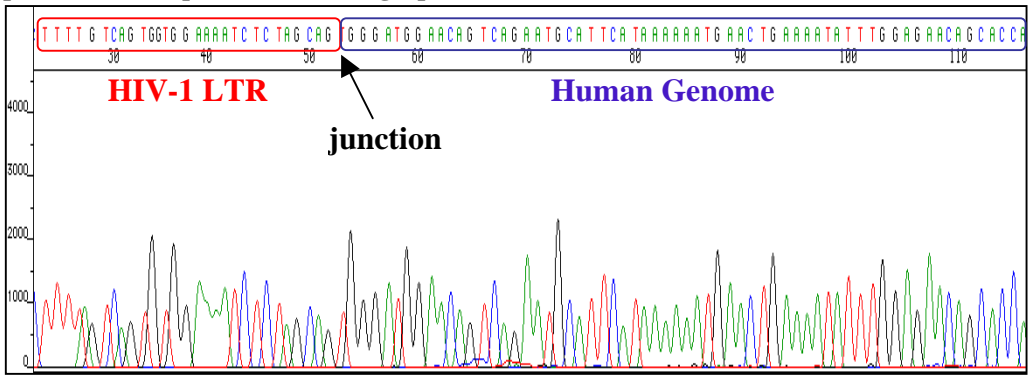

An example of a sequencing chromatograph showing a unique single sequence for the integration site. The HIV-1 LTR junction with the human genome is shown just above the chromatograph sequence.

**GenomeWalker PCR and Sequencing Primers**

The GenomeWalker protocol utilizes two sets of Gene Specific Primers (GSP). GSP1 is for the 1° PCR reaction, and GSP2 used for the nested 2° PCR reaction.

In designing primers, we used BLAST to ensure no significant homology existed with sequences in the human genome and performed GenomeWalker PCR on naïve Jurkats to ensure that no regions were preferentially amplified.

Schematics of all primers used in relation to the HIV LTR are presented below. Primers listed in the figures but whose sequences are not provided failed to yield PCR products or generated non-specific amplification.

For amplifying genomic sequence upstream of the HIV-1 5' LTR U3 region:
(these primers are anti-sense to a region just downstream of the 5' U5)
GSP1-1: TTCAGCAAGCCGAGTCCTGCGTCGAGA
GSP2-2: TCCCTTTCGCTTTCAAGTCCCTGTTCG

For amplifying genomic sequence downstream of the HIV-1 3' LTR U5 region:
(these primers are sense to a region just upstream of the 3' U3)
U3GSP1-1: GGTGGGTTTTCCAGTCACACCTCAGGT
U3GSP2-1: CCTTTAAGACCAATGACTTACAAGGCA

For amplifying genomic sequence downstream of the HIV-1 3' LTR:
(these primers are sense to a region just upstream of the 3' U3, and could NOT be used with the *Dra* I library)
3'GSP1-1: TCTGAGCCTGGGAGCTCTCTGGCTAAC
3'GSP2-1: TAGGGAACCCACTGCTTAAGCCTCAAT

For amplifying genomic sequence downstream of the HIV-1 3' LTR U5 region:
(these primers are sense to a region within R/U5)
3'GSP1-U5: TCTCTCTGGTTAGACCAGATCTGAGCCT
3'GSP2-U5: CCTCAATAAAGCTTGCCTTGAGTGCTT
OR
3'GSP1-U5b: TCTCTGGCTAACTAGGGAACCCACTGC
3'GSP2-U5b: AAGCCTCAATAAAGCTTGCCTTGAGTG

Sequencing primers used were:
U3'as: CCTTCTCTTGCTCAACTGGT (anti-sense within U3)
3'U5for: CCTCAATAAAGCTTGCCTTGAGTGCTT (sense at R/U5)
3'U5for2: CAAGTAGTGTGTGCCCGTCT (sense within U5)
*For sequencing the LTR to detect mutations:*
U3'for: ACTCCCAACGAAGACAAGAT (sense at 5' edge of U3)
U3'for2: ACTGCTGATATCGAGCTTGC (sense at 3' edge of U3)

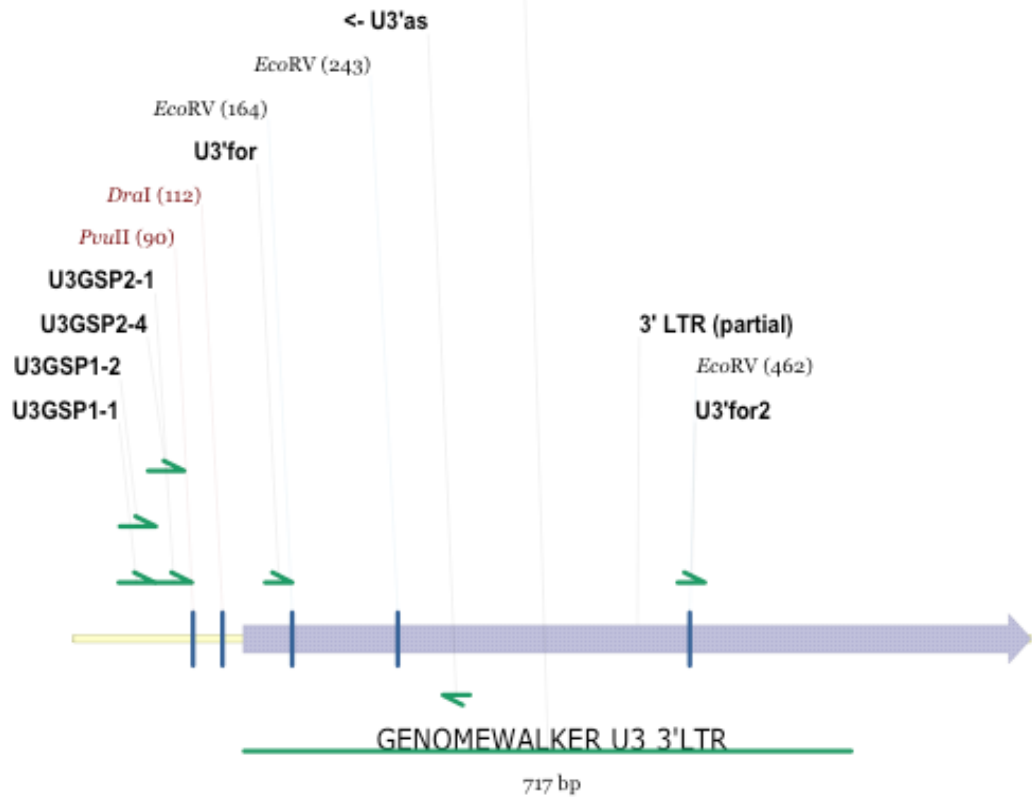
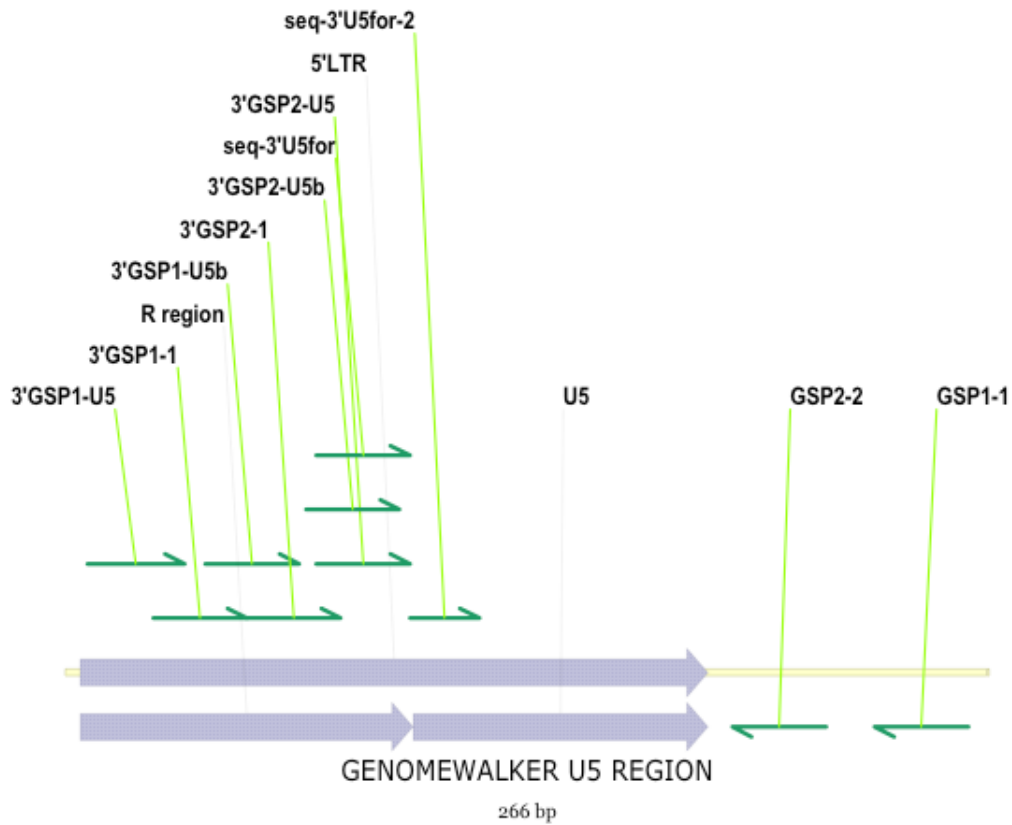

## 2. Supplemental Gene Expression Dynamics Figures

The relative proportions of different phenotypes obtained from a clonal populations isolated from a *Dim* sort of the LGIT infection (see Fig. 2, manuscript):

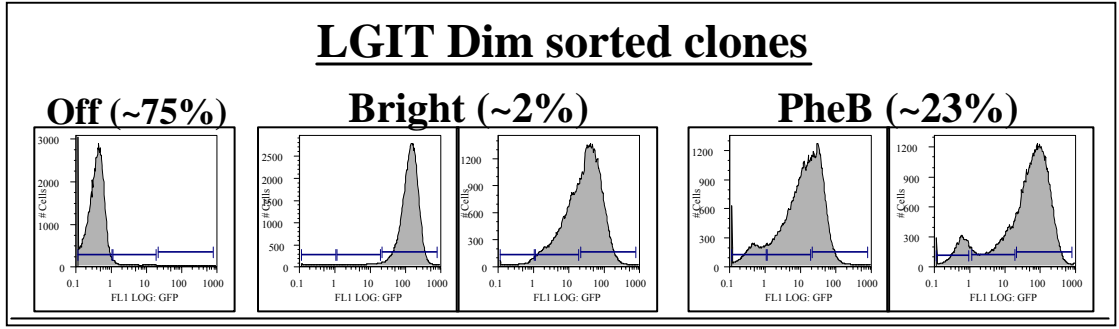

### Chemical Perturbations of a Representative LTR-GFP (LG) *Dim* Sorted Clone

The original unperturbed LG *Dim* clone is in red. The LG clone behavior is shown after 17 hour incubation in: TNFα (green), PMA (orange), and PMA+TNFα (cyan) and after infection with a retroviral vector expressing Tat from a strong tetracycline transactivator regulated promoter (blue). The retroviral vector, CLPIT Tat, is a variant of a virus previously described (Ignowski and Schaffer, 2004).

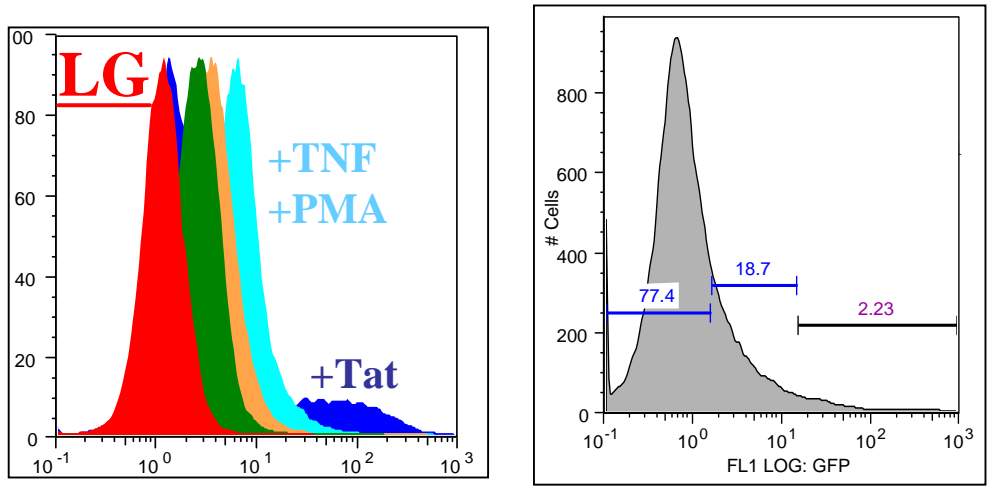

### LTR-GFP-CMV-PURO[R] Control Infection to Measure Integration Bias

To verify that LG integration was not biased to chromosomal regions of high LTR/GFP expression, Jurkat cells were also infected with LTR-GFP-CMV[P]-Puro[R] (LGCP) where Puro[R] was the Puromycin N-acetyl-transferase antibiotic resistance gene under the control of the Cytomegalovirus immediate early promoter (CMV[P]). Before antibiotic selection the flow cytometry profile of LGCP infected culture appeared identical to that of an LG infection. After antibiotic selection, the vast majority of LGCP infected cells exhibited no GFP expression, but GFP expression could be activated to Dim/Mid levels in all cells by stimulation with the LTR upregulator tumor necrosis factor alpha (TNFα) (data not shown). <u>The results indicated no integration bias to regions of high basal rate and suggested that many integration sites have little or no basal transcription</u> in agreement with (Jordan et al. 2001). The LGCP infection results demonstrate that only ~20% of integrations have a detectable basal rate.

LGCP was cloned by inserting the CMV-*pac* cassette into LG between the *Xho* I and *Bam* HI sites. LGCP was a kind gift from Josh Leonard.

**Dynamics of GFP Fluorescence after LGIT Infection of Jurkat and 293 Cells**

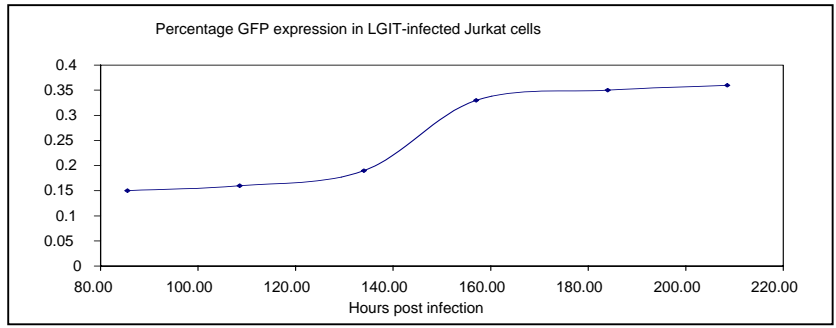

Legend: As described in the Results section of the main article, ~500,000 Jurkat or 293 cells were infected with purified LGIT virus (typically $10^7$-$10^8$ infectious units/ml, as determined by PMA+TNF$\alpha$ activation after infection) and analyzed by flow cytometry. Time = 0 corresponds to the infection time. 100,000 cells were analyzed by flow cytometry at each time point.

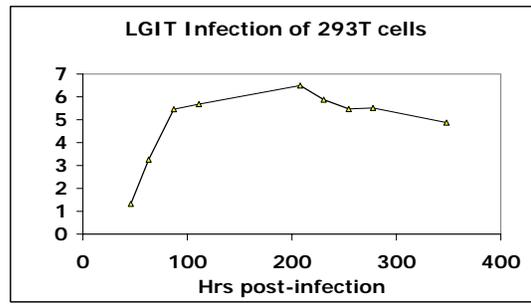

Jurkat and 293 kidney epithelial cells were infected at a low multiplicity of infection (MOI) with the LG and LGIT lentiviral vectors. Typically, after infection with retro- or lentiviral vectors expressing GFP under the control of a standard Cytomegalovirus immediate early promoter ($CMV^P$), the expression profile stabilizes within 2-4 days (Jordan et al. 2001; Ignowski and Schaffer 2004). Likewise, LG-infected Jurkat cells as well as LGIT-infected 293 cells stabilized in fluorescence profile after ~3 days, as previously reported (Jordan et al. 2001). In contrast, LGIT-infected Jurkats stabilized after only ~6 days. This delay was presumably due to a lower rate of Tat translation, mediated by the IRES, which increased the time necessary to accumulate a threshold concentration of Tat for transactivation.

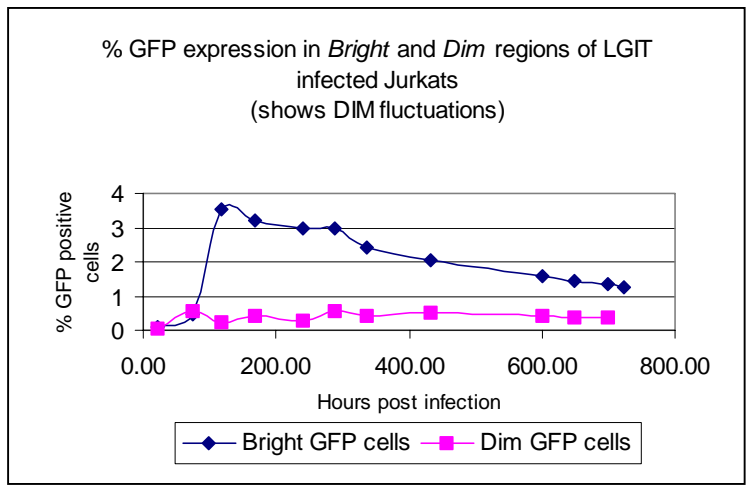

**Extrinsic Noise Controls**

We have argued that the phenotypic bifurcation we observe is due to noise intrinsic to the Tat feedback loop, rather than sources of noise that affect the level of gene expression or fluorescence of the cell as a whole. We have therefore systematically performed a number of key controls for extrinsic noise sources that collectively emphasize the importance of intrinsic Tat feedback noise.

*i. Cell Cycle*

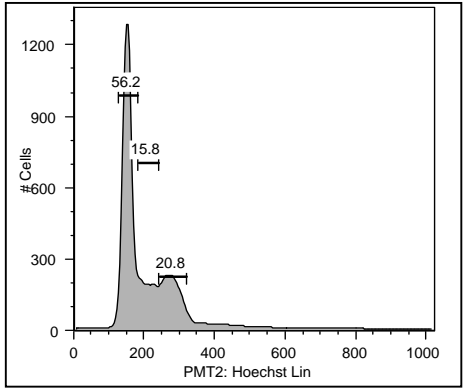

Flow cytometry analysis of Hoescht (Molecular Probes, Eugene OR) DNA staining of an LGIT PheB clone. 3 gates are visible within this histogram: G1 (56.2%), S (15.8%), and G2 (20.8%).
Cells were gated from each region, and GFP fluorescence was examined.
No bias for GFP expression or PheB was found in any of the cell cycle states, cells from each state of the cell cycle displayed PheB.

Flow cytometry profiles of GFP Fluorescence in various LGIT PheB cell cycle states

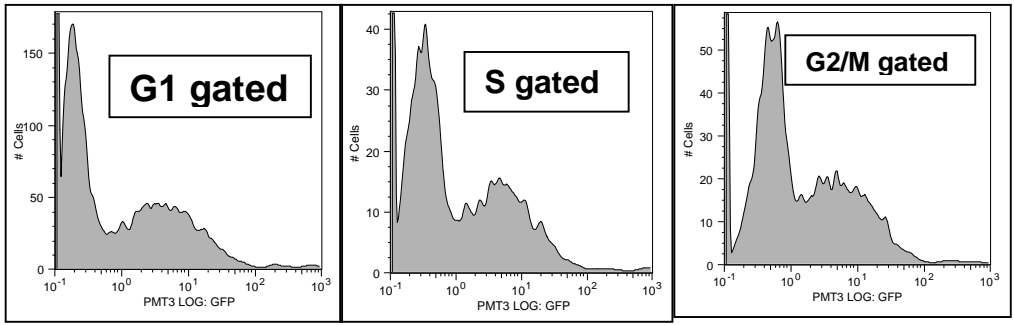

*ii. Cell size and Aneuploidy*

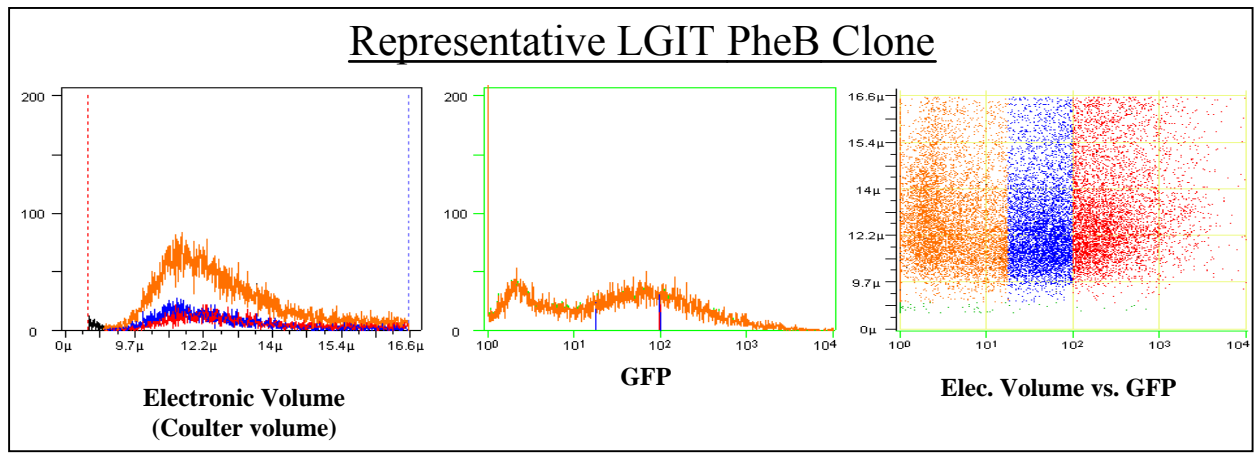

PheB clones were analyzed on a NPE Quanta flow cytometer (see Methods, main text). Fixation in 2% formaldehyde had no effect on volume measurement or fluorescence. The left panel shows that all fluorescence subpopulations, *Dim/Mid* (blue) and Bright (red), have the same cell volume distribution as the overall population (orange). The middle panel shows GFP fluorescence. Right panel (volume vs. GFP) clearly shows no bias in cell volume with respect to GFP fluorescence in PheB.

### *iii. Testing Tat Secretion by Coculture of LG Dim and LGIT PheB Jurkat Clones*

We and others have demonstrated that when exogenous Tat protein is incubated with cells, it is capable of trafficking to the nucleus and modulating HIV gene expression (Jordan et al. 2001). If in addition Tat were secreted from cells, it could potentially be taken up and modulate the gene expression of neighboring cells, and this higher level feedback could complicate the interpretation of experiments. We therefore explored and eliminated this possibility. Cells were counted, and equal numbers of cells were mixed and analyzed by flow cytometry at indicated time points. If Tat were being secreted and transactivating nearby cells, then some LG *Dim* cells should be transactivated, and the percentage of cells in the Bright region should increase from 31.9%. This did not occur. Turning *Off* of Bright cells (as seen previously) appear to be accounting for the increase in the *Off* percentage.

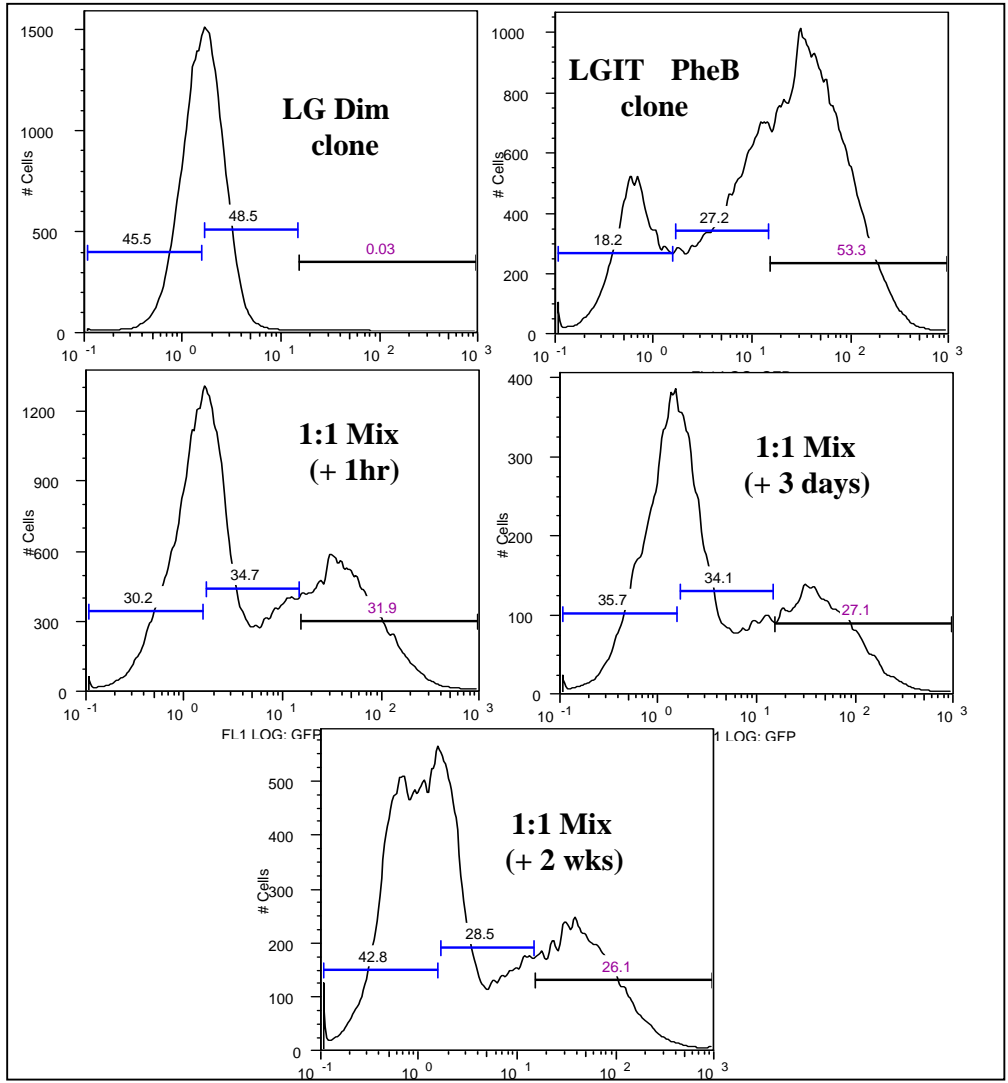

*iv.    Mitosis*

Another possible explanation for unequal GFP expression between cells within a clonal population is that GFP is unequally segregated during cell division.  We now present data eliminating this possibility.  Below is a representative micrograph of M phase enrichment (by nocodazole wash technique) of an LGIT PheB clonal population.  Over 300 cells were examined and no unequal GFP distribution was observed in Bright, *Dim*, or *Off* dividing cells. (Below is a dividing *Dim* cell:  Green is GFP, Red in CM-DiI (Molecular Probes, Eugene OR) staining of the plasma membrane, and blue is DAPI (Molecular Probes, Eugene OR) stained DNA.

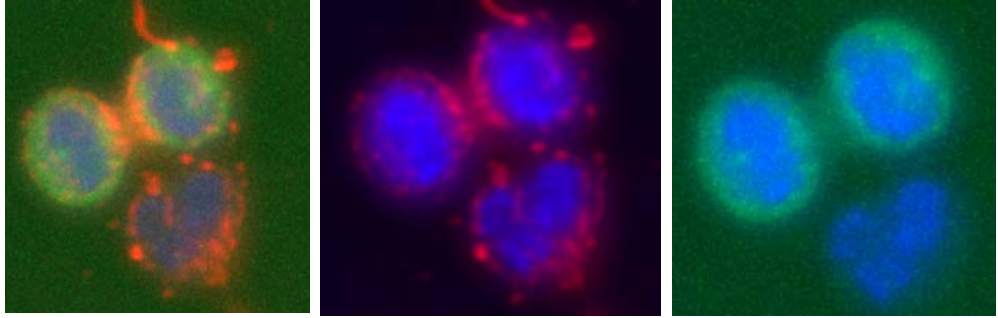

GFP segregation between daughter cells was only observed in one tenuous case in a non-M phase enriched population.  For completeness we have included this micrograph below and we invite the reader to make their own conclusions as to whether these cells are recent mitotic daughters.  Below is a Deconvolution Micrograph and 3D reconstruction (using Imaris Bitplane software) of a dividing LGIT PheB cell with unequal GFP distribution to daughters.  Red is cytoplasm, green is GFP, and blue is DAPI stained DNA.

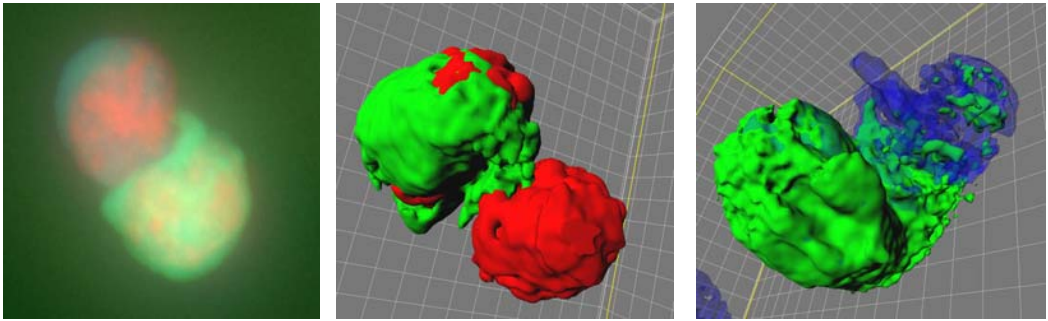

## 3. Supplemental Experimental Procedures
**Heparin Incubations**

Tat is known to bind the poly-sulfate form of the glyco-amino-glycan heparin (Tyagi et al. 2001), and the membrane bound form, heparan sulfate, has been implicated as the receptor utilized by Tat to cross the plasma membrane during protein transduction. A series of LGIT infections at different MOIs were performed, and soluble heparin (Sigma Chemical Co, St. Louis MO) added to the cultures shortly after infection to inhibit the cellular uptake of potentially secreted Tat. Infections were not performed in the presence of heparin since it would be sequestered by the polyvalent cation polybrene that is used to increase infection efficiency. If Tat were being emitted from GFP bright LGIT infected cells and transactivating LGIT *Dim* cells, we would expect to see a reduction the proportion of LGIT *Bright* cells upon heparin incubation. No reduction in *Bright* GFP expression was observed in these heparin-incubated cultures (see above). Furthermore, an LGIT-infected Jurkat clone expressing high levels of GFP (manuscript, Fig. 3d) was co-incubated with a low expressing LG clone (manuscript, Fig. 3f) that was shown to transactivate to under exogenous recombinant Tat protein incubation at a 1:1 ratio. If Tat were emitted from the LGIT infected cells the proportion of low GFP expressing cells (LG clone) should have decreased. But no change in the relative concentrations of each GFP population was observed (see above). The equivalent co-culture experiment was performed using bulk sorted (poly-clonal) populations of LGIT *Bright* and LG cells (supp info). These results agree with unpublished observations obtained by other groups in which they co-cultured Jurkats transfected with a Tat-expressing plasmid and Jurkats infected with an LTR-luciferase lentiviral vector and observed no luciferase expression (E. Verdin personal communication).

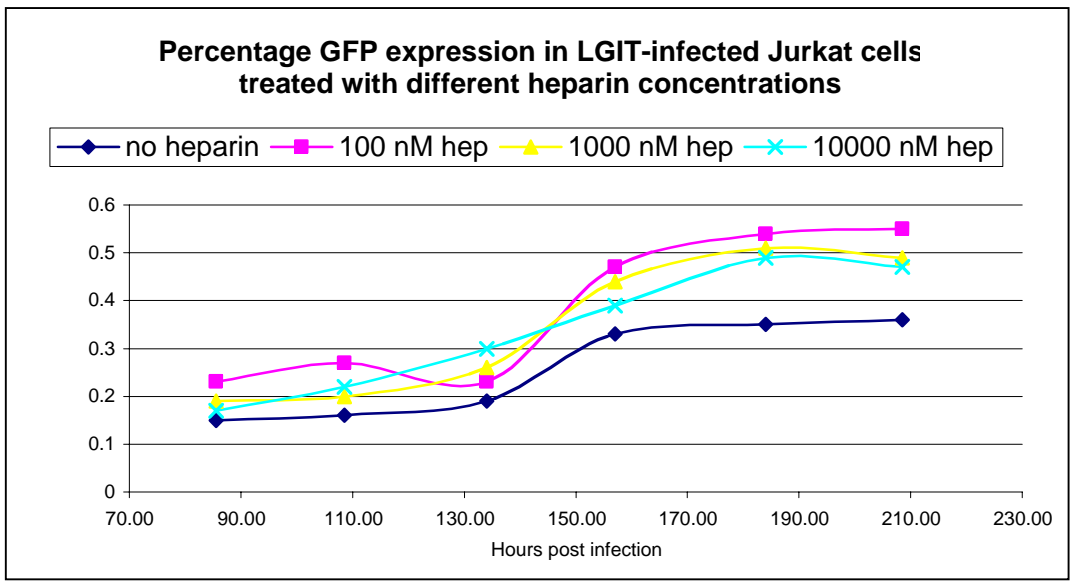

Similar heparin incubations were performed in 293T kidney epithelial cells infected with the LGIT vector and yielded similar results that did not support Tat trafficking between cells:

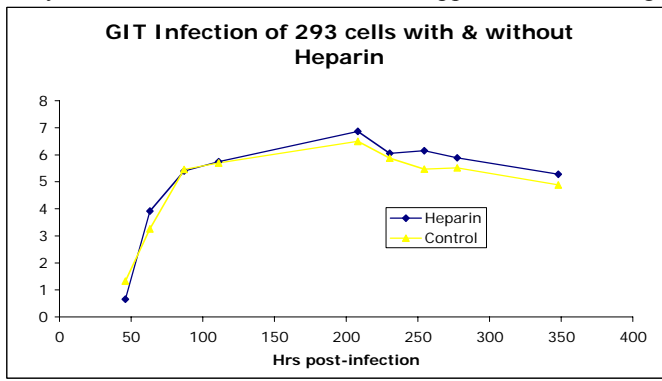

**In Silico Design of the In Vitro Stochastic Reporting Vector LGIT**

   We used published data (Feinberg et al. 1991; Reddy and Yin 1999) to construct a preliminary ODE model that could fit Tat transactivation dynamics observed in cell culture (Feinberg et al. 1991). The model described only the nuclear and cytoplasmic mRNA concentration of a reporter protein driven off the HIV-1 LTR in the presence of Tat as well as the concentration of this reporter protein. Nonlinear least squares fitting of the ODE model to published experimental data was performed using Berkeley Madonna in order to establish general parameter regimes for unknown or unpublished parameter values.

   In order to study this Tat transactivation model in the limit of low Tat concentration (a regime where noise could affect the output of the circuit) the ODE model was converted into a stochastic Monte-Carlo simulation via the Gillespie algorithm (Gillespie 1976). Stochastic simulations were performed using an in-house code written in ANSI C adapted from previous work (Lai et al. 2004). nRNA and cRNA are nuclear and cytoplasmic mRNA, respectively, P is protein (in this case the reporter Neo used by Feinberg et al. 1991).

**ODE model**

$d/dt \, (tat) = - dtt*tat$
$d/dt \, (nRNA) = (b+v*tat)/(k+tat) - ex*nRNA - dr*nRNA$
$d/dt \, (cRNA) = ex*nRNA - dr*cRNA$
$d/dt \, (P) = vp*cRNA/(kp+cRNA) - dp*P$
$RNA = nRNA + cRNA$
$dtt = 0.154; \ b = 0.01; \ dr = 1.6; \ ex = 2.6; \ dp = 0.39; \ v = 150; \ k = 50; \ vp = 11; \ kp = .676$

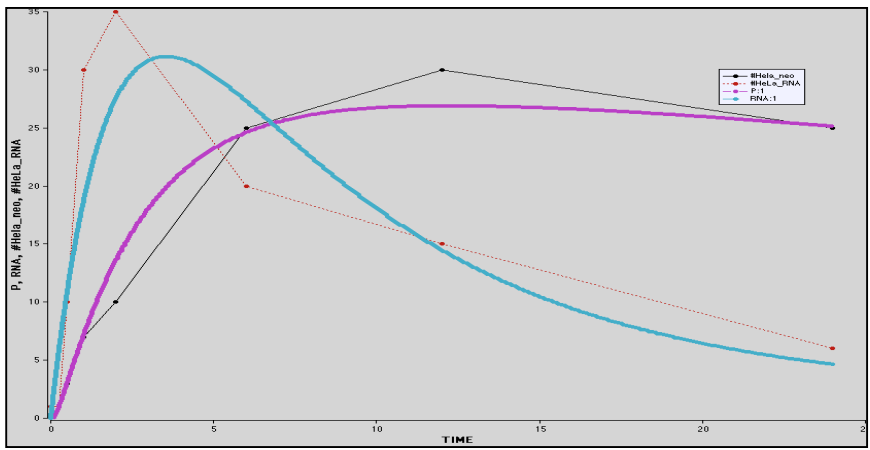

The model was numerically solved in Berkeley Madonna™ using published parameter values (Reddy and Yin 1999) and nonlinear least squares fitting *in vitro* data (Feinberg et al. 1991) for unknown parameter values. This ODE model was converted into a stochastic simulation by the method of Gillespie and simulations produced the following results, where Tat concentration as a function of time (seconds) is shown:

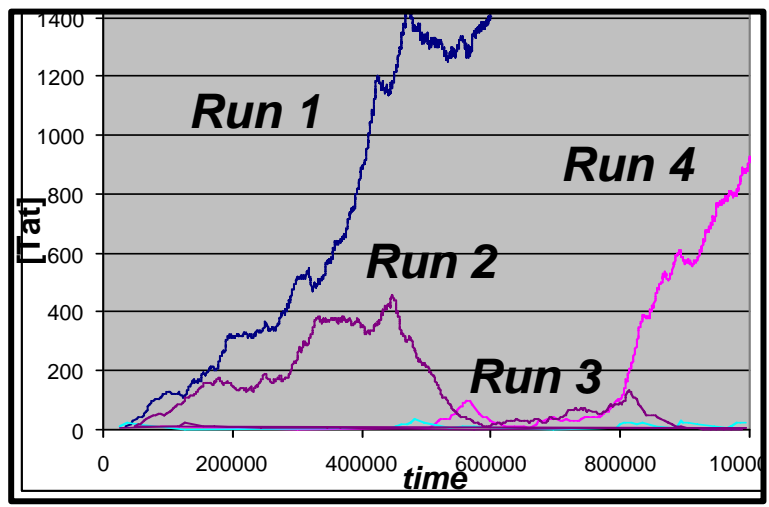

Outcome was highly dependant upon basal and Tat degradation rates, and produced a highly variable outcome where the system either transactivated (Run 1), began transactivation and then failed (Run 2), on occasion never transactivated (Run 3), or transactivated after a long delay (Run 4).

The LGIT vector was designed such that Tat followed the IRES in order to reduce the effective basal rate for Tat. GFP was used as a reporter for the activity of the HIV-1 long terminal repeat (LTR) promoter, and Tat was placed 3' of the Internal Ribosomal Entry Sequence (IRES) of encephalomyocarditis virus, since the IRES is known to reduce expression of the $2^{nd}$ cistron (i.e. Tat) 10-fold or more, relative to the first cistron (i.e. GFP) (Mizuguchi et al. 2000). Thus, the IRES was used to effectively reduce the Tat transcription rate to amplify the effects of the low Tat expression dynamics that we wished to study, somewhat analogous to Tat's position downstream of several splice acceptor options in wild type HIV-1. A fraction of wild type HIV-1 clones that integrate into regions of the genome nonpermissive for high basal gene expression would be expected to yield the same low expression rates. In order to isolate the effect of the Tat transactivation loop, no other HIV-1 genes were included in the vector.

**Sub/Resorting Experiments on LGIT PheB Clones**

LGIT PheB clones were routinely FACS sub-sorted into *Bright* and *Off* populations, and these sub-sorted populations consistently had initial fluorescence profiles in the region from which they were sorted. *Bright* sorted sub-populations did relax into the *Off* fluorescence (Davis et al. 2001) region over time. We also provided Jurkat LGIT PheB clones to the Laboratory of David Schatz's (Yale Univ. Medical School), and equivalent sub-sorting results and relaxation dynamics were obtained in their hands.

Cells sub-sorted from the *Dim* region of an LGIT PheB clone evolved and recapitulated the original bifurcated profile. In rare cases FACS sorting from the *Off* region of an LGIT PheB clone resulted in a fraction of cells turning *Bright*. This result may be due to the stress (both fluid dynamical and high-voltage stress) induced by FACS.

**No Correlation of Aneuploidy with GFP Expression using Flow Cytometry**

Thomas et al (Thomas et al. 2002) have reported a technique to measure aneuploidy in cell lines. Briefly, the NPE Systems flow cytometer is a unique device that gates fluorescence events based on volume (Coulter volume, or electronic impedance) rather than the standard forward vs. side-scattering gating method used by Coulter and Beckton-Dickinson cytometers. Accurate measurements of volume can be made for cells, nuclei, and beads based on the impedance change caused by volume exclusion in a solution of known ionic strength. This principle along with DAPI staining of DNA can be used to measure aneuploidy in cells. Typically, DAPI staining in this system is used for cell cycle analysis, as a cell of course doubles the amount of DNA in the nucleus between G1 and G2, and the cell and its nucleus are known to expand in volume from G1 to S to G2. (This volume expansion phenomenon is the basis of a cell cycle state isolation technique called centrifugal elutriation (Davis et al. 2001)). Thus, nuclei of the same size but containing more DNA are likely polyploid, and this has been confirmed in transformed breast cancer epithelial cell culture lines (NPE Systems, personal communication). The excess nuclear volume does not appear to be due to be taken up by excess space or water (as determined by deconvolution microscopy, see US Patent # 4,818,103).

Jurkat cells were fixed in 2% formaldehyde, washed in PBS, and resuspended in NIM-DAPI solution (NPE Systems Inc., Pembroke Pines, FL) to lyse the membrane. Non-lysed, formaldehyde-fixed cells exhibited volume measurements equivalent to non-formaldehyde fixed cells (~15 μm diameter). GFP-expressing Jurkat cells lysed in NIM-DAPI, but not fixed in formaldehyde (i.e. nuclei only) exhibited volume measurements equivalent to Jurkat cells lysed in Triton X-100 (i.e. nuclei only, ~5 μm diameter) and did not exhibit any GFP fluorescence, consistent with the removal of all cytoplasmic proteins. Surprisingly, Jurkat cells that had been fixed and then lysed in NIM-DAPI registered volume measurements equivalent to bare nuclei (~5μm diameter) while still exhibiting GFP fluorescence equivalent to unfixed, non-lysed cells. This result can be explained by recalling that Coulter volume is measured by exclusion of ions in solution and lysing + fixation still allows for ions to permeate the fixed cytoplasm but not the nucleus. Thus, fixed + lysed cells maintain a halo of GFP around the major volume determinant, the nucleus, and using this protocol it is possible to correlate aneuploidy with GFP expression. As can be seen in the figure below, for two representative LGIT PheB clones, no significant aneuploidy was observed (Elec. Vol. vs DAPI), and no correlation could be observed between *Bright* or *Off* subpopulations and DNA content vs. nuclear volume (i.e. aneuploidy), respectively.

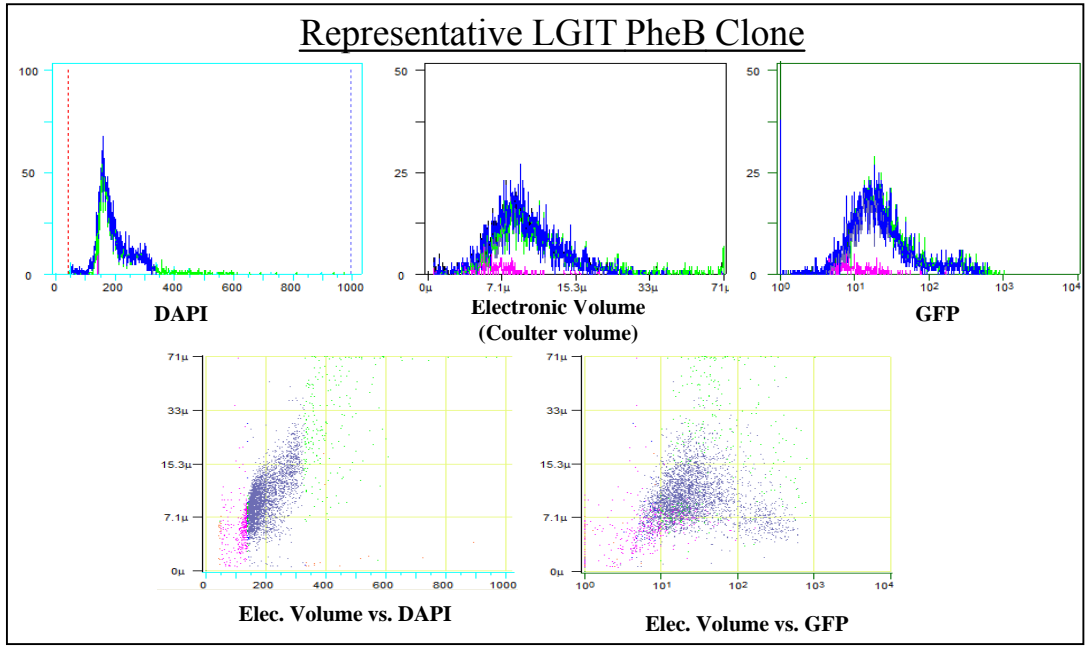

Furthermore, nuclear volume did not appear to be correlated with increased nuclear protein levels, as determined by fluorescamine staining using the above fixation + lysing protocol (data not shown).

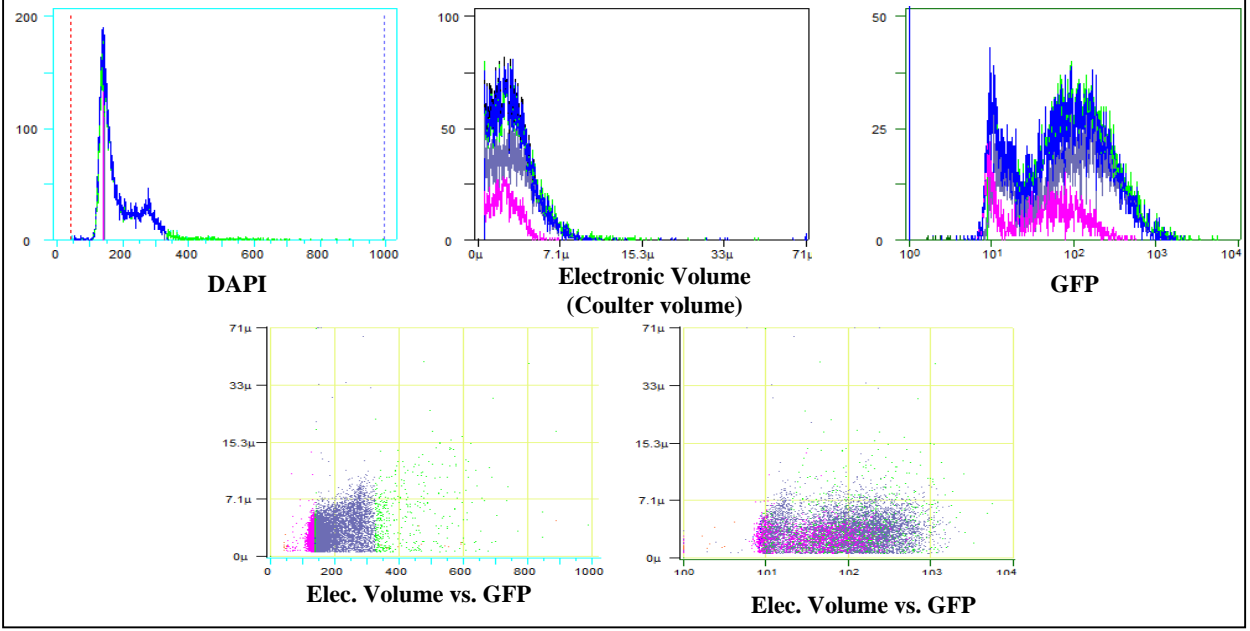

### Increased Transactivation Efficiency by Downregulation of the pTEFb Inhibitor HEXIM1

HEXIM1 binds to the pTEFb complex, specifically Cyclin T1 and CDk9, via the small nuclear RNA 7SK and renders pTEFb inactive (Yik et al. 2004). In order to titrate away HEXIM1 and thereby upregulate pTEFb we constructed a retroviral vector over-expressing 7SK under the control of an RNA polymerase III promoter (U6 promoter) and encoding an antibiotic selection cassette (CMV-Neo$^R$). LGIT Bright bulk sorts (Fig. 2a) and individual PheB clones (Fig. 3b) were infected with this vector and selected using neomycin. As shown (left), LGIT cells over-expressing 7SK were uniformly *Bright*, had slightly increased *Bright* fluorescence compared to naïve LGIT, and did not relax into the *Off* region over many weeks. A control vector where 7SK was replaced by an inert GFP sequence (which does not appear to generate protein when driven from the RNAPII promoter) showed no increased fluorescence and did relax into the *Off* region.

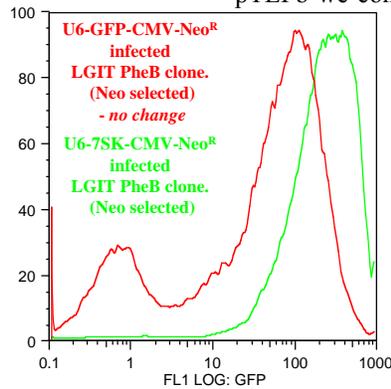

### LTR-mRFP-IRES-TatGFP Infection

The monomeric Red Fluorescent Protein (mRFP) (Campbell et al. 2002) was chosen over conventional dsRed for its rapid maturation time (on same the order of GFP) and its inability to generate fluorescence resonance energy transfer (FRET) with GFP. MOI was estimated based upon tittering virus after infection with naïve Jurkat cells followed by incubation in TNFα, as described in the main text. The difference in mean GFP fluorescence between LGIT and LRITG after TNFα incubation provided a rough/preliminary measure of IRES translation efficiency between the 1$^{st}$ and 2$^{nd}$ cistron. LGIT GFP fluorescence was ~200 fold greater than LRITG GFP fluorescence after TNFα stimulation (data not shown), though the GFP-Tat fusion may have a different protein stability and translation efficiency as compared to GFP alone.

### qRT-PCR Analysis of LGIT and Nearby Gene Expression

Clones were sorted with the Beckman-Coulter EPICS Elite ESP Sorter into *Bright* and *Off* collections of 500,000 cells. Total RNA from unsorted clones and each sorted collection was immediately isolated using TRIzol reagent (Invitrogen, Carlsbad, CA) and quantified by spectrophotometry. Then, total RNA was reverse-transcribed to cDNA using the ThermoScript RT-PCR System (Invitrogen) with oligo(dT)$_{20}$ primers according to the manufacturer's instructions. Each RT-PCR reaction was performed at 55.0$^o$C for 50 min. Quantitative real-time PCR (qPCR) was performed using the iCycler iQ Real-Time PCR Detection System (Bio-Rad, Hercules, CA) and SYBR Green or Black Hole Quencher (BHQ) probes. SYBR green was used for all qPCR reactions, except for the detection of HIV Ψ, in which the previously reported Taqman probe (Biosearch Technologies, Novato, CA) [5' 6-FAM d(AGCTCTCTCGACGCAGGACTCGGC) BHQ-1 3'] was employed (Sastry et al. 2002). For all reactions, qPCR conditions were: 95.0$^o$C for 2 min followed 55 cycles of 95.0$^o$C for 30 s, an annealing temperature for 30 s, and 72.0$^o$C for 20 s. The annealing temperature was 67.0$^o$C for human beta-actin, 60$^o$C for HIV Ψ, 57.0$^o$C for human LAT1-3TM, and 55.0$^o$C for both human FOXK2 and human C11orf23. Oligonucleotide primer sequences (Invitrogen) are as follows:
Sense human beta-actin (5'- ACCTGACTGACTACCTCATGAAGATCCTCACCGA), antisense human beta-actin: (5'-GGAGCTGGAAGCAGCCGTGGC CATCTCTTGCTCGAA), sense human LAT1-3TM (5'-CTCAAGCCGCT CTTCCCCA), antisense human LAT1-3TM (5'-GGCCTTCACGCTGTAGCAGTTCA) (Ito et al. 2002), sense HIV Ψ (5'-ACCTGAAAGCGAAAGGGAAAC), antisense HIV Ψ (5'-CACCCATCTCTCTCCTTCTAGCC) (Sastry et al. 2002), and the RT$^2$PCR Primer Sets for human FOXK2 (a.k.a. ILF1) and human C11orf23 (SuperArray Bioscience Corporation, Frederick, MD). Data were normalized to human beta-actin, and all reactions were perfomed in triplicate. To confirm the qPCR specificity, melt curves were performed on the Bio-Rad iCycler for all samples assayed with SYBR Green, and all samples were resolved on a 3% agarose gel to confirm the presence of a specific band.

Below are the raw data that was used to construct Fig. 4d in the main text.

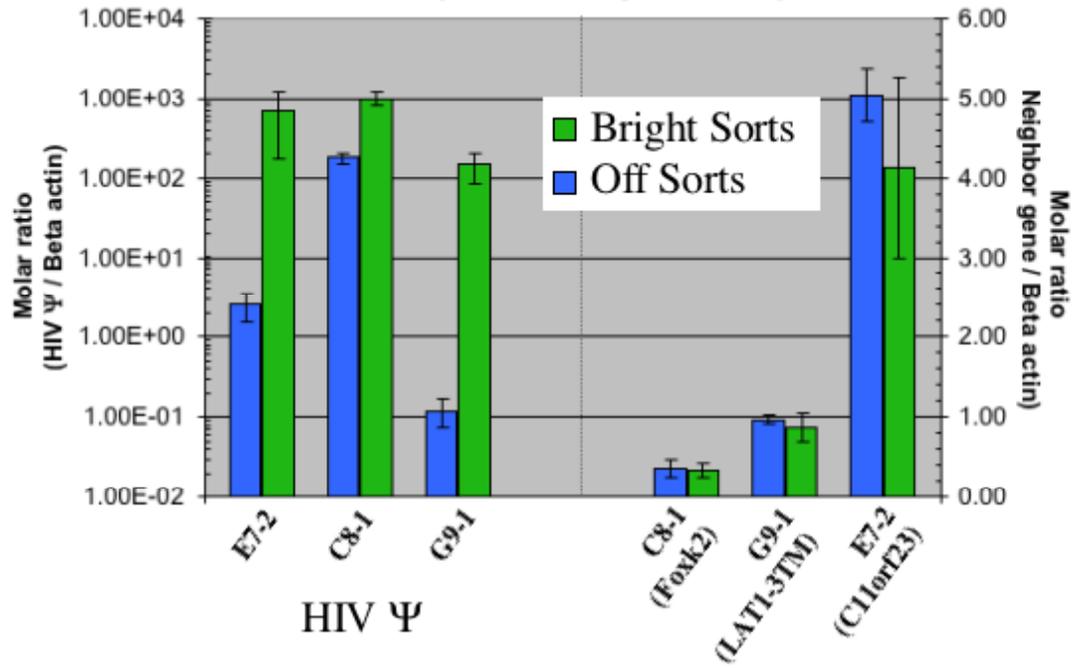

## 4. Summary Table of Extrinsic Controls
Here we present summaries of many of the controls performed, the potential mechanism being controlled for, and the result of the experiment:

Noise Extrinisic to Tat Transactivation Is Not Responsible for PheB

| Argument | IRES induces noise in protein translation and results in GFP diff. between cells. | Cell size: larger cells have >GFP fluor. than smaller cells. | Cell cycle state produces GFP differences in cells | Mitosis produces GFP diffs. in daughter cells |
|---|---|---|---|---|
| Evidence against | - IRES Literature does not support IRES amplifying translational noise<br>- Previous results with LTIG (Jordan et al. and our own) do not show IRES inducing GFP instability. | Measurements on the NPE flow cytometer that measures true Coulter volume (not light scattering) showed no cell size bias in diff. fluorescence regions | Flow cytometery cell cycle analysis of LGIT PheB clones (DAPI staining) showed no cell cycle bias for different regions of fluorescence | Nocodazole wash experiment showed no GFP differences between mitotic daughter cells in >300 cells observed. |

LGIT PheB Clones Are a Monoclonal Population without Significant Poly- or Aneuploidy

| Argument | Errors in FACS single-cell sorting distributed 2 or more cells into each well producing a polyclonal pop. | LGIT PheB Jurkat clones are aneuploid with *Bright* cells having multiple copies of chromosomes where LGIT integrated |
|---|---|---|
| Evidence against | - Genomewalker PCR showed single bands for all PheB clones on both 5' and 3' amplification<br>- Genomewalker using 2 clonal pops. Mixed together produced 2 bands and polyclonal bulk FACS sort produced multiple bands (Fig. 5)<br>- Sequencing of total PCR product produced a single sequence | - Flow cytometry experiment (see methods below) to simultaneously measure aneuploidy (DAPI) and GFP content in a cell showed no bias for different fluorescence subpopulations. |

FACS Sorting from the *Dim* Region Is Not Inherently Unstable

| Argument | FACS sorter has poor fidelity near *Dim* region |
|---|---|
| Evidence against | LG *Dim* bulk FACS sort was stable and did not relax into different fluorescent regions, and LG *Dim* clonal FACS sorts did not exhibit PheB. |

LGIT PheB *Off* Subpopulation Does Not Consist of Transcriptionally "Dead" Cells (Incapable of Supporting GFP Expression/Fluorescence) or Cells with a Mutation in *gfp*, *tat*, or the HIV LTR

| Argument | *Off* cells in LGIT PheB clones are transcriptionally dead |
|---|---|
| Evidence against | Chemical perturbations (Fig. 4) using TNF, PMA, or TSA induce *Off* cells in PheB clones to become *Bright*. Incubation in exogenous Tat protein does the same. |

PheB Is Not a Result of Growth Rate Selection; *Bright* and *Off* Populations Do Not Double at Different Rates

| Argument | *Bright* & *Off* cells grow at different rates, thereby producing 2 different populations |
|---|---|
| Evidence against | - growth curves of FACS sorted *Off* and *Bright* subpopulations from an LGIT PheB clone were identical (repeated for 2 different PheB clones)<br>- increasing transactivation rate (via titration of the transactivation inhibitor HEMIM1 by constitutive overexpression of 7SK) turns most cells *Bright* and the remaining *Off* population does not increase in proportion over many weeks. |

Stochastics (aka Molecular Fluctuations) in Tat Are Relevant Even at Low Expression Levels

| Argument | Stochastics or molecular fluctuations brought about by low levels of Tat are not relevant, the experiments are not necessarily consistent with stochastic theory |
|---|---|
| Evidence against | Higher levels of Tat expression show deterministic behavior:<br>- LGIT *Bright* FACS clonal sorts did not exhibit PheB<br>- LTIG (Tat before IRES, thus higher translation rate) bulk FACS sorts (*Dim*, *Mid*, *Bright*) did not exhibit relaxation.<br>- LTIG *Dim* and *Mid* clonal FACS sorts did not exhibit PheB<br>- 293 cells (known to have high NF-κb activation, equivalent to TNFα stimulation) infected with LGIT did not exhibit relaxation upon bulk FACS sorting and did not exhibit PheB upon *Dim* FACS clonal sorting.<br>- 7SK overexpression reduces stochastics by biasing forward loop |

### Tat Emission and Subsequent Uptake by Other Cells Does Not Contribute to PheB

| | |
|---|---|
| Argument | Tat emission from some LGIT infected cells with high basal rate may transactivate other nearby LGIT cells with a low basal rate. |
| Evidence against | - Highly controversial mechanism in HIV research field, other groups have been unable to reproduce Tat emission from infected cells (Karn 2000).<br>- Incubation of LGIT infected Jurkat cells in soluble Heparin polysulfate (potent Tat binding agent) does not decrease proportion of GFP positive cells in culture.<br>- Clonal mixing of LGIT *Bright* and LG *Dim* cells in culture produces no transactivation of LG *Dim* clones. |

### PEV-like Mechanisms Cannot Account for PheB

| | |
|---|---|
| Argument | PheB can be completely accounted for by PEV-like heterochromatin spreading from HERVs and stochastics are irrelevant. |
| Evidence against | - Only 7/17 PheB clones evidenced HERV proximate integrations.<br>- Little evidence that HERVs mediate heterochromatin spreading, and no evidence that isolated HERV LTRs do so.<br>- Exogenous Tat could transactivate PheB clones with HERV proximate integrations indicate that the HIV LTR was not inaccessible due to heterochromatin (Fig. 4c).<br>- Expression of genes nearest to (or overlapping with) LGIT PheB integration sites was not different between Bright and Off sorts under qPCR analysis (Fig. 4c).<br>- The chromatin remodeling agent Trichostatin A did not significantly alter the expression of any gene adjacent to or overlapping with an LGIT PheB integration site. |

## 5. Simulation Details

Stochastic Monte Carlo simulations were performed by direct simulation of the Chemical Master Equation according to the method of Gillespie (Gillespie 1976, 1977), using a speed amplification algorithm (Gibson and Bruck 2000); code is available upon request. The simulation code (adapted from (Lai et al. 2004)) was written in C++ and run on a Macintosh dual 2Ghz G5. GFP trajectories and histograms were plotted with PLPLOT (http://www.plplot.org). EGFP calibration beads (Clontech, Palo Alto, CA) were used to determine the molecules-to-RFU conversion: EGFP molecules = 37700*RFU – 4460 for our Beckmann Coulter cytometer. Histograms were generated using an algorithm similar to that employed by a flow cytometer: the RFU value (0.1 to 1000 units) was assigned to one of 1024 equally sized bins. All parameters in the model were varied over at least 2 orders of magnitude for reproducibility and sensitivity analysis (data not shown).

To simulate population dynamics after *Dim* and *Mid* bulk LGIT FACS, random trajectories falling within *Dim* and *Mid* GFP regions after 1 week of virtual simulation time (i.e. $\sim 6 \times 10^5$ seconds and corresponding to simulation-initiation/cell-infection) were tracked for up to 3 weeks (virtual time). To simulate clonal FACS of LGIT and reproduce PheB, a single random trajectory falling within the *Dim* GFP fluorescence region (1 virtual week after simulation-initiation/cell-infection) was tracked and recorded after $\sim 4$ weeks virtual time, or $\sim 2.4 \times 10^6$ seconds. *Dim* simulations were initiated with $Tat_0 = 5 \pm 4$ mols. and $GFP_0 \approx 75,000 \pm 50,000$ mols. in order to mimic the fidelity of bulk FACS sorting. *Mid* bulk sort simulations were initiated with $Tat_0 = 25 \pm 10$ mols. and $GFP_0 \approx 300,000 \pm 100,000$ mols.

For all simulations, average parameter values, in seconds$^{-1}$ ($s^{-1}$) and molecules (mols.), are as follows: $k_{BASAL} = 10^{-8}$ (transcripts/sec), $k_{EXPORT} = .00072$ ($s^{-1}$), $k1_{TRANSLATE} = .5$ ($s^{-1}$), $k2_{TRANSLATE} = .005$ ($s^{-1}$), $k_{BIND} = 10^{-4}$ (mols.$^{-1} \times s^{-1}$), $k_{UNBIND} = 10^{-2}$ ($s^{-1}$), $k_{ACETYL} = 10^{-3}$ (mols.$^{-1} \times s^{-1}$), $k_{DEACETYL} = 0.9$ ($s^{-1}$), $k_{TRANSACT} = 0.1$ ($s^{-1}$). GFP, Tat, & mRNA decay rates were taken from published values (Reddy and Yin 1999) while $k1_{TRANSLATE}$ vs. $k2_{TRANSLATE}$ values were determined from the 2-reporter LRITG control (above). All simulations were initialized with one LTR molecule ($LTR_0=1$), and all other species, except for Tat and GFP, set to zero molecules.

K50A simulations were performed by perturbing $k_{ACETYL}$ from $10^{-3}$ (mols.$^{-1} \times s^{-1}$) for wild-type LGIT to $0.0007$ (mols.$^{-1} \times s^{-1}$) for K50A LGIT mutant simulations. %Off trajectories for K50A (Fig. 5d, manuscript) were generated using a variant of the code above that calculated the percentage of trajectories in the *Off* region at each time point specified.

In cases where considering cell division was essential to directly compare simulations with experimental kinetics, parameters above were modified to account for a 24 hour Jurkat division time. For example, the GFP decay rate constant was increased so that when cell division was taken into account GFP "RFU" had a 16-hour half-life, as opposed to the true 48-hour GFP "molecular" half-life. This was done since dividing cells dilute GFP to mitotic daughters and thus, in a population of actively dividing cells, the decay rate of GFP+ cells appears much faster than the GFP molecular decay rate.

**Low Bright versus Very Bright Computational Prediction and Experimental Results**

Eqs. 1-13 (manuscript) made a number of predictions that were subsequently confirmed by experiment. Here we report upon the decay characteristics of trajectories sorted *in silico* from different areas of the *Bright* LGIT region. Since the *Bright* region is predicted to be a quasi-stable state or mode continually maintained by Tat positive feedback, different areas of the *Bright* region should decay to, and thus populate, the *Off* state at differing rates. Specifically, cells in the *Low-Bright* region (i.e. closer to the *Mid* region) should be less "stable" and decay to *Off* more quickly than cells from the *Mid-Bright* (i.e. the GFP peak in an LGIT infection) or *Very-Bright* regions (i.e. cells populating the very bright regions of *Bright*). Molecularly, *Low-Bright* cells may have less Tat, CDk9, or Cyclin T1 and thus have slower forward reaction rates steering the transactivation circuit towards the *Off* state.

Simulations of *Very-Bright* and *Low-Bright* sorts showed that *Low-Bright* sorts did populate the *Off* region more quickly than *Very-Bright* sorts. These model-based decay rate predictions were then confirmed by *in vitro* FACS analysis of cells sorted from the *Very-Bright* and *Low-Bright* regions of LGIT infected Jurkats (below).

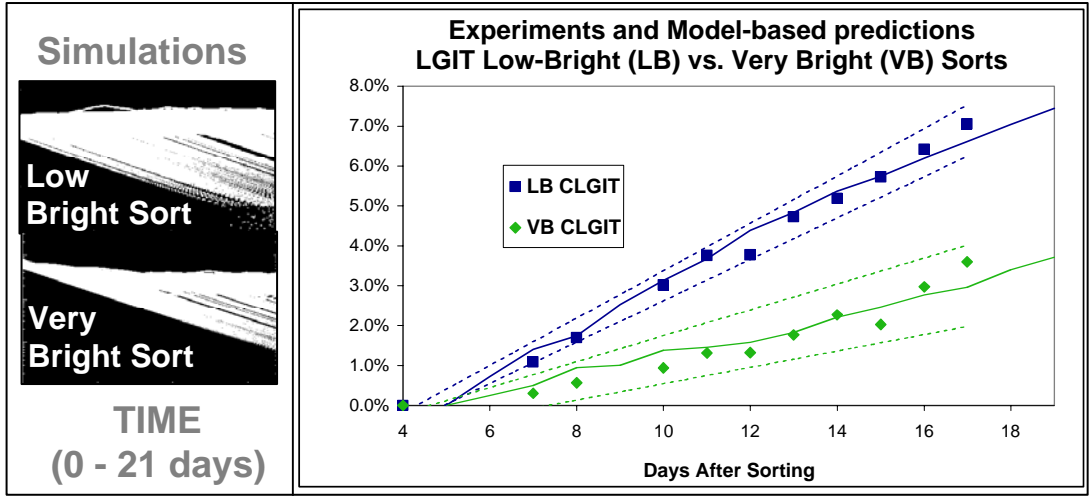

Legend: Left: Gillespie simulations of Eqs. 1-13 (manuscript) for *Very-Bright* and *Low-Bright* initial conditions. Right: Percentage of Jurkat cells in the *Off* region of after sorting from the *Very-Bright* (green diamonds) and *Low-Bright* (blue squares) regions of an LGIT infection. Dashed lines represent upper and lower 95% confidence intervals for the correspondingly colored data. Solid lines are the results of 10,000 trajectories (Eqs. 1-13) from *Very-Bright* (green) and *Low-Bright* (blue) simulations. *Mid-Bright* simulations and FACS analysis appeared indifferentiable from *Very-Bright* simulations and FACS sorting (data not shown).

# 6. Supplemental Discussion
**LGIT *Bright* Clones Do Not Exhibit PheB, in Agreement with Stochastic Theory**

Small random molecular fluctuations constantly occur in physical process, but their effect is typically significant only when a system contains few molecules or involves slow kinetic processes. When large numbers of molecules are present in the system, random molecular fluctuations typically have a negligible influence (except near bifurcation points or other distinguished dynamical points) and are effectively averaged out. This 'averaging out' phenomenon is observed when stochastic Monte Carlo simulations are run with large numbers of molecules (Gillespie 1977). Therefore, we performed control experiments to test whether increased levels of Tat expression eliminated PheB.

First, Jurkat clones FACS sorted from the *Bright* region of an LGIT infected Jurkat culture were found not to exhibit PheB (Fig. 3b). In order to rule out the possibility that the *Dim* region of GFP fluorescence was inherently unstable to FACS sorting, Jurkat clones were also FACS sorted from the *Dim* region of the LG infection, and these clones remained stable in the *Dim* region for many weeks (Fig. 3b)—an expected result given the absence of Tat or positive feedback in this system.

As an additional control we explored the dynamics of LTR-Tat-IRES-GFP (LTIG) infection of Jurkat cells (Jordan et al. 2001). The LTIG vector (a kind gift from Eric Verdin, UCSF) generates a higher level of Tat expression than LGIT, since Tat precedes the IRES, and should display reduced stochastic behavior. At low MOI, LTIG infection produced a *Mid* subpopulation that stabilized ~3 days post-infection, similar to the dynamics of the LG infection. LTIG-infected Jurkats did not form a *Bright* subpopulation—in agreement with (Jordan et al. 2001) and presumably due to reduced GFP translation caused by the GFP placement after the IRES—whereas bulk FACS sorted LTIG *Mid*, *Dim*, and *Off* subpopulations did not generate relaxation kinetics, and clones FACS sorted from the LTIG *Dim* region did not generate PheB (data not shown). These LTIG FACS results again support the hypothesis that a higher Tat expression level abrogates the effects of random molecular fluctuations in Tat.

As another control, 293 kidney epithelial cells were infected with LGIT at low MOI. 293 cells are known to have high levels of NF-κB activation (Horie et al. 1998), indicating that the HIV-1 LTR would be highly activated in these cells, and any integration site would exhibit high rates of basal Tat expression. 293 and Jurkat cells infected with LGIT appeared very similar under flow cytometry, with the exception that 1) extremely few cells (<0.1%) existed in the *Dim* or *Mid* regions of GFP fluorescence and 2) the *Bright* GFP subpopulation stabilized ~3.5 days after of infection rather than the 6-8 days in Jurkat cells (see below). Populations of cells and clones FACS sorted from the *Dim* and *Mid* regions of 293 LGIT infections did not exhibit relaxation kinetics or PheB (data not shown).

**Controls: Stochastic Processes Extrinsic to Transactivation Do Not Contribute to PheB**

In order to differentiate between stochastic molecular fluctuations intrinsic to the chemical reactions of Tat transactivation and noisy processes extrinsic to the Tat transactivation loop (e.g. cell cycle), an array of control experiments were performed to measure the contribution of these processes to PheB. No correlation could be found between PheB and cell cycle state, cell volume, or aneuploidy in LGIT PheB population (see below).

Unequal distribution of GFP to daughter cells upon mitotic division may also account for phenotypic clonal bifurcation, though this is unlikely due to the high number of GFP molecules (~10,000/cell, Supp. data) required to detect GFP fluorescence in our flow cytometry apparatus. In order to examine noise in mitotic division as a source of PheB, PheB clones were enriched for dividing, S phase, cells by nocodazole wash (Davis et al. 2001) and cells analyzed by microscopy. Over 300 mitotic cells of various different GFP expression levels were examined (data not shown). Cells could be found expressing *Bright*, *Dim*, and *Off* levels of GFP, but GFP segregation between daughter cells was only observed in one tenuous case (see above). Thus, unequal distribution of GFP to daughter cells upon mitotic division does not appear to be a significant mechanism accounting for PheB. Although this result does not preclude the possibility that PheB results from division mediated unequal distribution of Tat, mitosis seems unlikely to be a major factor during infection of $CD4^+$ T lymphocytes *in vivo* since HIV-1 kills these cells in under 2 days, and cells revert to memory in ~5-7 days (Pierson et al. 2000).

Stochastic mechanisms in cellular emission of Tat and subsequent uptake by neighboring cells could provide an alternate avenue of extrinsic stochastic noise that may lead to clonal phenotypic bifurcation. Extracellular HIV-1 Tat is known to 'transduce' the plasma and nuclear membranes and transactivate any HIV-1 LTRs present in the cell (Karn 2000). Tat has been reported to be is emitted from HIV-1 infected cells (Ensoli et al. 1994), but the phenomenon remains controversial in the HIV-1/AIDS research field

(Karn 2000). Nevertheless, a number of control experiments were performed to assess whether Tat could be emitted from some LGIT infected cells and taken up by other cells to transactivate the LTR in those recipient cells (see below). Co-culturing of LGIT PheB or *Bright* clones with LG *Dim* clones failed to transactivate the LG *Dim* clones, and LGIT infection in the presence of heparin (a potent Tat binding agent) failed to increase the proportion of GFP-expressing, transactivated cells. These results appear to agree with the results of other groups who failed to find evidence of Tat emission from infected cells (Karn 2000) (E. Verdin, personal communication).

We next examined the possibility that PheB was merely an outcome of cell growth rate differences induced by GFP or Tat expression. Tat has been reported to be cytotoxic at high concentrations (Mattson et al. 2002), and GFP expression can in some cases increase the metabolic load placed on a cell, causing it to proliferate more slowly (DVS, unpublished observations). Therefore, growth rates of LGIT PheB sub-populations sorted from regions of *Bright* and *Off* regions, respectively, were compared. No growth rate differences could be measured over a two week period (data not shown). Furthermore, increasing the Tat transactivation efficiency, by downregulating the pTEFb inhibitor HEXIM1 (Yik et al. 2004), abrogated PheB (all cells shifted to the *Bright* region). Importantly, the *Off* sub-population did not re-establish even after many weeks (see below), inconsistent with a growth rate selection mechanism.

DNA methylation did not appear to be a mechanism accounting for PheB since the methylation inhibitor 5-azacytidine had no effect on GFP expression in PheB LGIT clones (data not shown), in agreement with previously reported results (Jordan et al. 2001; Pion et al. 2003).

*Potential Limitations*

The 2 exon (101 amino acid) version of Tat was used in this study. Tat quickly reverts to the 1 exon (76 amino acid) form after relatively few passages in cell culture (Fields et al. 2001). Although the deletion of the $2^{nd}$ exon is not known to affect transactivation, it is possible that this deletion might explain a portion of the non-PheB LGIT *Dim* sorted clones that integrated into positions very similar to the PheB LGIT clones but did not transactivate under TNF$\alpha$. Regardless, this would not impact the phenomenon of PheB in a single clone.

The IRES element was incorporated into the stochastic reporting vector in order to reduce the amount of basal Tat expression, hopefully into the regime where stochastic fluctuations became significant. Although inserting the ECMV IRES into an HIV-1 vector may appear divorced from HIV-1 physiology, it is unclear to us whether the resulting LGIT vector is more or less divorced from wild-type HIV-1 than the LTIG vector used previously (Jordan et al. 2001) or an LTR-TatGFP fusion. Many factors analogous to our placement of Tat downstream of the IRES contribute to a relative reduction in Tat levels in wild-type HIV-1. For example, *tat* is located ~5.3 kb downstream of the LTR, and RNAPII may prematurely disassociate before reaching the *tat* gene. Additionally, there are ~15 splice transcriptional splice variants produced from the HIV-1 provirus, and Tat is translated from only one of these. The LGIT vector does contain a splice site, but the IRES mediated translation reduction of Tat may be compensating for this lack of additional splice sites. Interestingly, wild-type HIV-1 has been reported to harbor at least one IRES element in the *gag* region (Buck et al. 2001). Finally, the IRES mediated reduction in Tat basal expression rate may be regarded as equivalent to HIV-1 integration into a site less permissive for expression from the LTR.

The introduction of the IRES could potentially introduce an additional source of noise, especially in light of reports that translation is an inherently noisy process (McAdams and Arkin 1999). However, many groups have worked extensively with IRES containing vectors, and we and others observe a perfectly linear correlation between expression of the $1^{st}$ and $2^{nd}$ cistron (Figure 3h) (Martinez-Salas 1999; Mizuguchi et al. 2000; Hennecke et al. 2001). Furthermore, the LTIG vector (which also contains an IRES) did not exhibit any stochastic phenomena (either bi-stability or phenotypic clonal bifurcation) in our hands, or as previously reported (Jordan et al. 2001). In fact Jordan et al. (2003) were able to develop stable latent/*Off* cell lines using LTIG, indicating a stable relationship between GFP and Tat at high and low expression levels. Interestingly, unpublished observations from another group using a wild-type HIV-1 vector where Tat has been replaced by the Tet-on system (Verhoef et al. 2001; Berkhout et al. 2002) exhibited behavior similar to PheB (B. Berkhout, personal communication). Lastly, the 2 reporter LTR-mRFP-IRES-TatGFP control infection showed a clear correlation between mRFP and GFP (Supp. data), arguing against IRES mediated noise in translation inducing PheB.

Our results demonstrated that only 47% of the PheB clones integrated within genes, in contrast to a recent reports of >90% of *in vivo* latent clones integrating in genes (Han et al. 2004). However, as others have noted, it is difficult to reconcile these integration results with an earlier study from this group (Hermankova et al. 2003) showing that 99% of *in vivo* isolated latent CD4 cells were defective and failed to produce virus upon stimulation. The *Pst* I digest used by Han et al. may also bias the detection method since cleavage sites for this enzyme are under-represented in mammalian genomes and are found primarily 5' of genes (Lindsay and Bird 1987).

There exists a well-founded counter-argument to premature transcription: an excess of histones in the nucleus (Alberts 1994) could drive rapid chromatin (including nuc-1) packaging of any naked DNA and thus establish the basal transcription rate. However, RNAPII may bind to the HIV-1 cccDNA pre-integration complex (PIC) and either transcribe before integration or synthesize one transcript after integration but prior to nuc-1 binding (a single transcript could readily generate 5-50 Tat proteins by multiple ribosome recruitment to the transcript). This possibility is plausible, as the barrier-to-autointegration factor (BAF) is the only chromatin protein currently known to associate with the HIV-1 PIC (Mansharamani et al. 2003), and little is known about it. There is also a possibility that integrase or one of the many other PIC associated proteins (Fields et al. 2001) may sterically inhibit nuc-1 formation in the PIC, and/or for a short time after integration. Furthermore, the first mature chromatin elements deposited on newly synthesized DNA are H2A/H2B (Smith and Stillman 1991), and RNAPII is known to readily displace these proteins during transcription (Studitsky et al. 2004). Importantly, Tat protein is not incorporated into HIV-1 virions, thus negating alternative explanations for a pre-basal concentration of Tat.

HIV-1 appears to be preferentially integrate in or near genes (Schroder et al. 2002), but integrations can occur virtually anywhere in the genome including in dense heterochromatic regions near centromeres (Jordan et al. 2003). Our data show that PheB can result from integrations in many regions of the genome, but integrations within 1 kb of HERV LTRs appears sufficient for PheB. We speculate that HERV LTRs may have associated heterochromatin that could spread 1 kb and establish a very low basal rate for LGIT. Unfortunately we have no direct evidence of such HERV mediated heterochromatin spreading, and the only epigenetic phenomenon reported to correlate with HERV expression is DNA methylation (Januchowski et al. 2004), which does not affect the HIV LTR (Pion et al. 2003). Little is known about the chromatin environment surrounding HERV LTRs. Studies in yeast show that acetylated and de-acetylated chromatin can spread to surrounding genomic areas and influence yeast mating type (Rusche et al. 2003).

Phenotypic bifurcation due to stochastic fluctuations in a transcriptional feedback loops could be a strategy evolved in diverse lentiviral variants to yield integrants poised at a knife-edge between active replication and latency. It may therefore be interesting to examine latent reservoirs in SIV infected Rhesus Macaques (McChesney et al. 1998) in addition to the HTLV Tax protein.